\def\half{\frac{1}{2}}

\def\mb#1{\mbox{\boldmath{$#1$}}}

\documentclass[amsmath,amssymb,showkeys, showpacs]{revtex4}
\usepackage{graphicx}
\begin{document}
\pdfoutput=1 % Force arXiv robot to use pdflatex
% ---------------------------
\hspace*{4.5 in}CUQM-138\\
%\hspace*{4 in} catom [24 March 2011] \vspace*{0.4 in}
% ---------------------------
\title{Spectral characteristics for a spherically confined $-a/r + br^2$ potential}

\author{Richard L. Hall $^1$, Nasser Saad $^2$, and K. D. Sen $^3$}
\address{$^1$ Department of Mathematics and Statistics, Concordia University,
1455 de Maisonneuve Boulevard West, Montr\'eal,
Qu\'ebec, Canada H3G 1M8}\email{rhall@mathstat.concordia.ca}
\address{ $^2$ Department of Mathematics and Statistics,
University of Prince Edward Island, 550 University Avenue,
Charlottetown, PEI, Canada C1A 4P3.}\email{nsaad@upei.ca}
\address{$^3$ School of Chemistry, University  of Hyderabad 500046, India.}
\email{sensc@uohyd.ernet.in}

\begin{abstract}
We consider the analytical properties of the eigenspectrum generated by a class of
central potentials given by $V(r)=-a/r + br^2,$ $b>0.$ In particular,
scaling, monotonicity, and energy bounds are discussed. The
potential $V(r)$ is considered both in all space, and under the condition of spherical
confinement inside an impenetrable spherical boundary of radius $R$. With the aid of the asymptotic iteration method, several exact analytic results are obtained which exhibit the parametric dependence of energy on $a, b$, and $R$, under certain constraints.
More general spectral characteristics are identified by use of a
combination of analytical properties and accurate numerical
calculations of the energies, obtained by both the generalized pseudo-spectral method, and the asymptotic iteration method. The experimental
significance of the results for both the free and
confined potential $V(r)$ cases are discussed.
\end{abstract}

\maketitle
%%%%%%%%%%%%%%%%%%%%%%%%%%%%%%%%%%%%%%
\section{Introduction}\label{intro}
%%%%%%%%%%%%%%%%%%%%%%%%%%%%%%%%%%%%%%
\noindent The model for a hydrogen atom, HA, confined in an impenetrable sphere
of finite radius $R$ was originally introduced \cite{mdb} to
simulate the effect of high pressure on atomic static dipole
polarizability. Sommerfeld and Welker \cite{sw} formulated the
wave function solutions for this potential in terms of confluent hypergeometric
functions, and underlined the application of this model for the
prediction of the line spectrum originating from atomic hydrogen
in the outer atmosphere.  An algorithm for obtaining nearly exact
energy calculations for a spherically confined hydrogen atom has
been published \cite{cb}. On the other hand, regular~soft
confinement of the Coulombic systems has been developed by
superimposing Debye screening \cite {sp}. Such a confining
potential has been successful in explaining \cite{ss,cbd} the
shift in frequency of the x-ray spectral lines emitted by
laser-imploded plasmas in the limit of high plasma density, whereby
the effective potential assumes the form given by the Coulomb
plus oscillator potential. The harmonic potential can be
considered here as giving rise to the confinement of the Coulomb
system with soft boundary walls. A variety of
other model potentials leading to the confinement of electrons in
atoms and molecules have been proposed, in order to explain the
behavior of the novel artificial nanostructures, such as
quantum wires and quantum dots, atoms and molecules embedded
inside fullerenes, zeolites and liquid helium droplets, and, in addition,
to simulate the interior of a giant planet. A
comprehensive review covering of the development and applications
of confining model potentials has been recently published \cite
{aqc,rv10}. Under the confinement effect of an impenetrable
spherical cavity of radius $R$, the hydrogen atom and the isotropic
harmonic oscillator, IHO, potentials have been studied,
independently, and their spectral characteristics have been
analyzed \cite {vip,sen} in terms of useful quasi exact results.
In the following text, we shall denote the spherically confined
hydrogen atom as SCHA and the spherically confined
isotropic harmonic oscillator as SCIHO: in both cases, the
eigenstates are labelled as $(\nu,\ell), \nu=1,2,3, \cdots,
\ell=0,1,2 \cdots$, in terms of which the number of radial  nodes for a given $\ell$
becomes $\nu -\ell-1.$

\vskip0.1true in
\noindent For the \emph{free} Coulomb plus oscillator potential, a few
exploratory calculations have been reported earlier
\cite{fmc,castro,wilets,pv}. In view of the importance of the
conjoined Coulomb and harmonic oscillator potential, it useful to
study the general behavior of this potential under the confinement
due to an impenetrable spherical cavity, as a function of the radius $R$,
where the free state is represented by ${R \rightarrow} {\infty}
$. In this paper we consider a general spherically symmetric model
of atomic system confined by (i) the presence of a
harmonic-oscillator potential term and in a representative set of
cases also (ii) containment inside an impenetrable spherical box
of radius $R$. In atomic units $\hbar = m = e = 1$ the Hamiltonian
for the model system is given by
\begin{equation}\label{hamitonian}
H = -\half \Delta + V(r),\quad V(r) = -\frac{a}{r} + b r^2,
\end{equation}
where $a$ and $b$ are coupling parameters. We shall always assume
that $b > 0$ and for the  most part, we shall assume that the
Coulomb term is also attractive, $a > 0;$ we shall also consider
the repulsive case $a < 0$,  in which $H$ becomes a model, for
example, for a system composed of a pair of confined electrons.
\vskip0.1true in
\noindent We shall now present a brief review of the known results defining
the spectral characteristics of the two confined systems SCHA and
SCIHO. It is well known that the so called \emph{accidental
degeneracy} of free HA is removed in the SCHA. As $R \rightarrow
0$, the energy levels $E(\nu,\ell)$ increase in magnitude such
that the higher $\ell$ states get relatively less destabilized.
There exists a critical value of $R$ above which $E(\nu,\ell) > 0$.
Further, two additional kinds of degeneracies arise \cite{vip}.
They result from the specific choice of the radius of confinement
$R$, chosen exactly at the radial nodes corresponding to the free
HA wave functions. In the \emph{incidental degeneracy} case, a
given confined $(\nu=\ell+1,\ell)$ state becomes iso-energic with
$(\nu=\ell+2,\ell)$ state of the \emph{free} HA with energy
$-1/\{2(\ell+2)^2\}$ a.u., at the same $R$.~In the
\emph{simultaneous degeneracy} case, on the other hand, a certain
pair of confined states at the common radius of confinement $R$
that is prescribed in terms of the location of the radial node in
a specific free state of HA, become iso-energic. For example, for
all $\nu \ge \ell+2$, each $(\nu,\ell)$ SCHA state is degenerate
with $(\nu+1,\ell+2)$ state, when both of them are confined at
$R=(\ell+1)(\ell+2)$, which defines the radial node in the free
$(\ell+2,\ell)$ state. Both these degeneracies have been shown
\cite{vip} to result from the Gauss relationship applied at a
unique $R$ by the confluent hypergeometric functions that describe
the general solutions of the SCHA problem.
\vskip0.1true in
\noindent We note that free IHO energy levels show the well-known
``$(2\nu+\ell)$'' degeneracy with the equidistant eigenvalues given
by $(2\nu+\ell-\frac{1}{2})\hbar \omega, ~\nu=1,2,3,\cdots, $~for
a given $\ell$. Such a degeneracy is removed under the confined
conditions. As $E(\nu,\ell) > 0$ at all $R$, the critical radius is
absent. The \emph{incidental degeneracy} observed in the case of
SCIHO is qualitatively similar to that of the SCHA. However, the
behavior of the two confined states at a common radius of
confinement is found to be different \cite{sen,spm}. In
particular, for the SCIHO the pairs of the confined states defined
by $(\nu=\ell+1,\ell)$ and $(\nu=\ell+2,\ell+2)$ at the common
$R=\sqrt{(2\ell+3)/2}$ a.u., display for all $\nu$, a constant
energy separation of \emph{exactly} $2$ harmonic-oscillator
units,~$2\hbar \omega$ , with the state of higher $\ell $
corresponding to lower energy. The choice of $R$ is qualitatively
similar to that in the case of SCHA, namely, it is the location of
the radial node in the $(\nu=\ell+1,\ell)$ state which is the
first excited state corresponding to a given $\ell$ for the free
IHO. It is interesting to note that the two confined states at the
common $R$ with $\Delta \ell=2$, considered above, contain different numbers of radial nodes.
\vskip0.1true in

\noindent With this background, we shall now consider the spherically
confined potential defined in Eq.(\ref{hamitonian}). The paper is organized as follows.
 In section 2, the scaling properties and monotonicity of the eigenspectrum
 generated by the potential $V(r)$, as a function of the parameters of the
 potential, are derived. Analytic energy bounds, derived by the envelope method, are
 reported in section~3: these are found to be useful in guiding the search for very accurate values by numerical methods. In sections~4 and 5, we use the asymptotic iteration method (AIM) to study
how the eigenvalues depend on the potential parameters $\{a,b,R\}$, repectively for the free system ($R = \infty$), and for finite $R$. In each of these sections, the results obtained are of two types: exact analytic results that are valid when certain parametric constraints are satisfied, and accurate numerical values for arbitrary sets of potential parameters. In section~6 we adjoin some more numerical data, obtained by the generalized pseudo-spectral (GPS) Legendre method, and present a detailed analysis of the spectral characteristics of the system and their experimental significance.

%%%%%%%%%%%%%%%%%%%%%%%%%%%%%%%%%%%%%%%%%%%%%%%%%%%%
\section{Scaling and monotonicity}\label{scaling}
%%%%%%%%%%%%%%%%%%%%%%%%%%%%%%%%%%%%%%%%%%%%%%%%%%%%
\noindent Since the potential and the confining box are spherically symmetric, we may write the energy eigenfunctions in the form
\begin{equation}\label{wavefunction}
\Psi(\mb{r}) = \frac{\psi(r)}{r}Y_{\ell}^{m}(\theta, \phi), \quad \psi(0) = 0,
\end{equation}
where $ \mb{r}\in\Re^3$ and $r = |\mb{r}|$. For finite box sizes $R$ we also require $\psi(R) = 0.$ In terms of the atomic units used, each discrete eigenvalue  depends on three  parameters. We shall express this by writing $E_{\nu\ell} = E(a, b, R)$. If we now introduce a scale factor (dilation) $\sigma >0 $  into the terms of the Hamiltonian,
so that $r \rightarrow \sigma r$, then, after multiplying the eigenequation
 $H\psi = E\psi$ through by $\sigma^2,$ we may derive the general scaling law
\begin{equation}\label{gscale}
E(a, b, R) =  \sigma^{-2}E(\sigma a,\, \sigma^4b ,\, R/\sigma), \, \sigma > 0.
\end{equation}
For example, the particular choices $\sigma = a^{-1}$, $\sigma = b^{-\frac{1}{4}}$, and $\sigma = R,$ then yield,
respectively, the special scaling laws
\begin{equation}\label{sscale}
E(a,b,R) = a^2E(1, b a^{-4}, aR) = b^{\half}E(a b^{-\frac{1}{4}}, 1,b^{\frac{1}{4}}R) = R^{-2}E(aR, bR^4,1).
\end{equation}
Thus it would be sufficient to consider just two spectral parameters.
\vskip0.1true in
\noindent The eigenvalues $E_{\nu,\ell} = E(a,b,R)$ are monotonic in each parameter. For $a$ and $b$, this is a direct consequence of the monotonicity of the potential $V$ in these parameters. Indeed, since $\partial V/\partial a = -1/r <0$ and $\partial V/\partial b = r^2 > 0$, it follows that
\begin{equation}\label{monotoneEab}
\frac{\partial E(a,b,R)}{\partial a} < 0\quad {\rm and}\quad\frac{\partial E(a,b,R)}{\partial b} > 0.
\end{equation}
The monotonicity with respect to the  box size $R$ may be  proved by a variational argument.
We shall show in section~(\ref{bounds}) that the Hamiltonian $H$ is bounded   below. The eigenvalues of $H$ may therefore be characterized variationally.  Let us consider two box sizes, $R_1 < R_2$ and an angular momentum subspace labelled by a fixed $\ell.$
We extend the domains of the wave functions in the finite-dimensional subspace spanned by the first $N$ radial eigenfunctions for $R = R_1$ so that the new space $W$ may be used to study the case $R = R_2$. We do this by defining the extended eigenfunctions so that $\psi_i(r) = 0$  for $R_1 \le r\le R_2.$  We now look at $H$ in $W$ with box size $R_2$.  The minima of the energy matrix $[(\psi_i,H\psi_j)]$ are the exact eigenvalues for $R_1$ and, by the Rayleigh-Ritz principle, these values are one-by-one upper bounds to the eigenvalues for $R_2.$ Thus, by  formal argument we deduce what is perhaps intuitively clear, that the eigenvalues increase as $R$ is decreased, that is to say
\begin{equation}\label{montoneER}
\frac{\partial E(a,b,R)}{\partial R} < 0.
\end{equation}
From a classical point of view, this Heisenberg-uncertainty effect is perhaps counter intuitive: if we try to squeeze the electron into the Coulomb well by reducing $R$, the reverse happens; eventually, the eigenvalues become positive and arbitrarily large, and less and less affected by the presence of the Coulomb singularity.
\vskip0.1true in
\noindent For some of our results we shall consider the system unconstrained by a spherical box, that is to say $R = \infty.$ For these cases, we shall write $E_{\nu\ell} = E(a,b).$  If a very special box is now considered, whose size $R$ coincides with any radial node of the $R=\infty$ problem, then the two problems share an eigenvalue exactly. This is an example of a very general relation which exists between constrained and unconstrained eigensystems, and, indeed, also between two constrained systems with different box sizes.

%%%%%%%%%%%%%%%%%%%%%%%%%%%%%%%%%%%%%%%%%%%%%%%%%%%%%%
\section{Some analytical energy bounds}\label{bounds}
%%%%%%%%%%%%%%%%%%%%%%%%%%%%%%%%%%%%%%%%%%%%%%%%%%%%%%
\noindent The generalized Heisenberg uncertainty relation  may be expressed \cite{RS2,GS} as the operator inequality $-\Delta > 1/(4r^2).$ This allows us to construct the following lower energy bound
\begin{equation}\label{lbound}
E > {\mathcal E} = \min_{0 <r \le R}\left[\frac{1}{8r^2} - \frac{a}{r} + b r^2\right].
\end{equation}
Provided $b \ge 0,$ this lower bound is finite for all $a$.  It also obeys the same scaling and  monotonicity laws as $E$ itself. But the bound is weak. For potentials such as $V(r)$ that satisfy $\frac{d}{dr}(r^2\frac{dV}{dr}) > 0,$  Common has shown \cite{common} for the ground state that $\langle-\Delta\rangle > \langle 1/(2r^2)\rangle,$ but the resulting energy lower bound  is still weak.
\vskip0.1true in
\noindent For the unconstrained case $R = \infty$, however, envelope methods
\cite{env1,env2,env3,env4,env5,envcp} allow one to construct analytical upper and lower energy bounds with general forms similar to (\ref{lbound}). In this case we shall write $E_{\nu\ell} = E(a,b).$  Upper and lower bounds on the eigenvalues are based on the geometrical fact that $V(r)$ is at once a concave function $V(r) = g^{(1)}(r^2)$ of $r^2$ and a convex function $V(r) = g^{(2)}(-1/r)$ of $-1/r$.  Thus tangents to the $g$ functions are either shifted scaled oscillators above $V(r)$, or shifted scaled atoms below $V(r)$. The resulting energy-bound formulas are given by
\begin{equation}\label{ebounds}
\min_{r > 0}\left[\frac{1}{2 r^2} -\frac{a}{P_1r} + b (P_1r)^2\right]\, \le \,E_{\nu\ell}(a,b)\,\le\, \min_{r > 0}\left[\frac{1}{2 r^2} -\frac{a}{P_2r} + b (P_2r)^2\right],
\end{equation}
where (Ref. \cite{env6} Eq.(4.4))
\begin{equation}\label{P12}
P_1 = \nu, \quad P_2 = 2\nu -(\ell+\half).
\end{equation}
We use the convention of atomic physics in which, even for non-Coulombic central potentials, a principal quantum number $\nu$ is used and defined
by $\nu = n+ \ell + 1,$ where $n$ is the   number of nodes in the radial wave function.  It is clear that the lower energy bound has the Coulombic degeneracies, and the upper bound those of the harmonic oscillator. These bounds are very helpful as a guide when we seek very accurate numerical estimates for these eigenvalues.
\vskip0.1true in
\noindent Another related estimate is given by the `sum approximation' \cite{env5}
 which is more accurate than (\ref{ebounds}) and is known to be a lower energy bound for the bottom $E_{\ell+1\,\ell}$ of each angular-momentum sub space: in terms of the $P$'s we have for these states, $P_2 = \nu + \half = P_1 + \half$.  The estimate is given by
\begin{equation}\label{sbound}
E_{\nu\ell}(a,b) \approx {\mathcal E}_{\nu\ell}(a,b) = \min_{r > 0}\left[\frac{1}{2 r^2} - \frac{a}{P_1r} + b (P_2r)^2\right].
\end{equation}
This energy formula has the attractive spectral interpolation property that it is {\it exact} whenever $a$ or $b$ is zero.
The energy bounds (\ref{ebounds}) and (\ref{sbound}) obey the same scaling and monotonicity laws is those of $E_{\nu\ell}(a,b).$  Because of their simplicity they allow one to extract analytical properties of the eigenvalues.  For example, we can estimate the critical oscillator coupling $\hat{b}$ that will lead to vanishing energy $E = 0.$  We may estimate $\hat{b}$ by using (\ref{ebounds}) or (\ref{sbound}).  We differentiate with respect to $r$, and use the vanishing of this derivative and of $E$ to obtain the following explicit formula for $\hat{b}$
\begin{equation}\label{critb}
\hat{b} \approx \left(\frac{27}{32}\right)\frac{a^4}{P_a^4P_b^2},
\end{equation}
in which $P_a$ and $P_b$ are to be chosen.  If $P_a = P_1$ and $P_b = P_2,$ then from (\ref{sbound}) we obtain a good general approximation for $\hat{b}.$  We can also obtain bounds on $\hat{b}$.
Since $E(a,b)$ is a monotone increasing function of $b$, we can state the nature of the bounds on $\hat{b}$ given by formula (\ref{critb}):
(i) if $P_a = P_b = P_1 = \nu,$ the formula yields an upper bound; (ii) if $P_a = P_b = P_2 = 2\nu -(\ell + \half),$ then it is a lower bound;
 (iii) if $\nu = \ell + 1$ and $P_a = \nu$ and $P_b = \nu + \half$, then the formula yields a lower bound.  We shall state this last result explicitly: for the bottom of each angular-momentum subspace, where $\nu = \ell + 1,$ the critical oscillator coupling $\hat{b}$ yielding $E = 0$ is bounded by
\begin{equation}\label{critb1}
\hat{b} \ge \left(\frac{27}{32}\right)\frac{a^4}{\nu^4(\nu+\half)^2}.
\end{equation}
%%%%%%%%%%%%%%%%%%%%%%%%%%%%%%%%%%%%%%%%%%%%%%%%%%%%%%%%%%%%%%%%%%%%%
\section{Exact solutions for the potential $V(r)$}\label{spec}
%%%%%%%%%%%%%%%%%%%%%%%%%%%%%%%%%%%%%%%%%%%%%%%%%%%%%%%%%%%%%%%%%%%%%
\noindent The radial three-dimensional Schr\"odinger equation for the Coulomb plus harmonic-oscillator potential, expressed in atomic units, is given by
\begin{equation}\label{eq13}
-{1\over 2}{d^2\psi(r)\over dr^2}+\left[{l(l+1)\over 2r^2}-{a\over r}+br^2\right]\psi(r)=E\psi(r), \quad 0<r<\infty,\quad b>0,~~ a\in \mathbb R
\end{equation}
where $l(l+1)$ represents the eigenvalue of the square of the angular-momentum
operator $L^2$. Note that for $a=0$, the potential $V(r)=-a/r+br^2$ corresponds to the pure harmonic oscillator potential, while for $a>0$, it is a sum of two potentials, the attractive Coulomb term $-a/r$ plus the harmonic-oscillator potential $br^2$. For $a<0$, the potential $V(r)$ corresponds to the sum of two potentials, the repulsive Coulomb potential $|a|/r$ plus a harmonic-oscillator potential $b r^2$.
\vskip0.1true in
\noindent Since the harmonic oscillator potential dominates at large $r$, this suggests the following Ansatz for the wave function:
\begin{equation}\label{eq14}
\psi(r)=r^{l+1}\exp(-\alpha r^2)f(r),
\end{equation}
where $\alpha$ is a positive parameter to be determined. Substituting this wave function into Schr\"odinger's equation (\ref{eq13}), we obtain the following second-order differential equation for $f(r)$:
\begin{equation}\label{eq15}
rf''(r)+(-4\alpha r^2+2l+2)f'(r)+((-2b+4\alpha^2)r^3+(-4\alpha l+2E-6\alpha)r+2a)f(r)=0,
\end{equation}
which suggest the value $\alpha=\sqrt{b/2}$. With this value of $\alpha$, Eq.(\ref{eq15}) is reduced to
\begin{equation}\label{eq16}
rf''(r)+\left(-2 r^2\sqrt{2b}+2(l+1)\right)f'(r)+\left[\left(2E-(2l+3)\sqrt{2b}\right)r+2a\right]f(r)=0.
\end{equation}
In order to find the polynomials solutions $f(r)=\sum_{k=0}^n a_k r^k$ of this equation, we rely on the following theorem (\cite{hakan1}, Theorem 5) that characterizes the polynomial solutions of a class of differential equations given by
\begin{equation}\label{eq17}
(a_{3,0}x^3+a_{3,1}x^2+a_{3,2}x+a_{3,3})~y^{\prime \prime}+(a_{2,0}x^2+a_{2,1}x+a_{2,2})~y'-(\tau_{1,0} x+\tau_{1,1})~y=0,
\end{equation}
where $a_{3,i},i=0,1,2,3$, $a_{2,j},j=0,1,2$ and $\tau_{1,k},k=0,1$ are arbitrary constant parameters.
\vskip0.1true in
\noindent{\bf Theorem 1.} \emph{
The second-order linear differential equation (\ref{eq17})
has a polynomial solution of degree $n$ if
\begin{equation}\label{eq18}
\tau_{1,0}=n(n-1)a_{3,0}+na_{2,0},\quad n=0,1,2,\dots,
\end{equation}
along with the vanishing of $(n+1)\times(n+1)$-determinant $\Delta_{n+1}$ given by
\begin{center}
$\Delta_{n+1}$~~=~~\begin{tabular}{|lllllll|}
 $\beta_0~~$ & $\alpha_1$ &$\eta_1$&~& ~&~ &~\\
  $\gamma_1$ & $\beta_1$ &  $\alpha_2$&$\eta_2$&~&~&~ \\
~ & $\gamma_2$  & $\beta_2$&$\alpha_3$&$\eta_3$&~&~\\
$~$&~&$\ddots$&$\ddots$&$\ddots$&~&~\\
~&~&~&$\gamma_{n-2}$&$\beta_{n-2}$&$\alpha_{n-1}$&$\eta_{n-1}$\\
~&~&~&&~$\gamma_{n-1}$&$\beta_{n-1}$&$\alpha_n$\\
~&~&~&~&$~$&$\gamma_{n}$&$\beta_n$\\
\end{tabular}~~=~~0
\end{center}
where its entries are expressed in terms of the parameters of Eq.(\ref{eq17}) by
\begin{align}\label{eq19}
\beta_n&=\tau_{1,1}-n((n-1)a_{3,1}+a_{2,1})\notag\\
\alpha_n&=-n((n-1)a_{3,2}+a_{2,2})\notag\\
\gamma_n&=\tau_{1,0}-(n-1)((n-2)a_{3,0}+a_{2,0})\notag\\
\eta_n&=-n(n+1)a_{3,3}
\end{align}
Here, $\tau_{1,0}$ is fixed by Eq.(\ref{eq18}) for a given value of $n$; the degree of the polynomial solution.
}
\vskip0.1true in

\noindent Consequently, for the polynomial solutions of Eq.(\ref{eq16}), we must have, by means of Eq.(\ref{eq18}), that
\begin{equation}\label{eq20}
E_{nl}=(n+l+{3\over 2})\sqrt{2b}
\end{equation}
and the conditions on the potential parameters are determined by the vanishing of the tri-diagonal determinant with entries
\begin{align}\label{eq21}
\beta_n&=-2a\notag\\
\alpha_n&=-n(n+2l+1)\notag\\
\gamma_n&=2(n-k-1)\sqrt{2b}\notag\\
\eta_n&=0
\end{align}
namely, the vanishing of the $(n+1)\times (n+1)$-tridiagonal determinant
\begin{center}
$\Delta_{n+1}$~~=~~\begin{tabular}{|c c c c c c |}
 $-2a~~$ & $-(2+2l)$ &$~$&~& ~&~\\
  $-2k\sqrt{2b}$ & $-2a$ &  $-2(3+2l)$&$~$&~&~\\
~ & $2(1-k)\sqrt{2b}$  & $-2a$&$-3(4+2l)$&$~$&~\\
& & & & &  \\
~&$~$&$\ddots$&$\ddots$&$\ddots$&~\\
& & & & &  \\
~&~&$2(n-3-k)\sqrt{2b}$&$-2a$&$-(n-1)(n+2l)$& \\
~&~&~&$2(n-2-k)\sqrt{2b}$&$-2a$&$-n(n+2l+1)$\\
~&~&$~$& & $2(n-k-1)\sqrt{2b}$&$-2a$\\
\end{tabular}
\end{center}
\vskip0.1true in
\noindent For $n=0$ we have, for the purely harmonic oscillator $a=0$, the exact energy
\begin{equation}\label{eq22}
E_{0l}= (l+{3\over 2})\sqrt{2b}
\end{equation}
which gives the ground-state $f_0(x)=1$ in each subspace labelled by the angular momentum quantum number $l$.
\vskip0.1true in
\noindent For $n=1$, the determinant $\Delta_2=0$ forces the potential parameters $a$ and $b$ to satisfy the equality
\begin{equation}\label{eq23}
a^2-{\sqrt{2b}(l+1)}=0
\end{equation}
with a necessary condition for the eigenenergy
\begin{equation}\label{eq24}
E_{1l}=(l+{5\over 2})\sqrt{2b}.
\end{equation}
The condition (\ref{eq23}) gives two possibilities for the wavefunction solution. First, for $a=-\sqrt{\sqrt{2b}(l+1)}$, i.e. with repulsive Coulomb term, we have a ground-state (no-node) eigenfunction given by
\begin{equation}\label{eq25}
\psi_0(r)=r^{l+1}\exp(-\sqrt{b\over 2}~ r^2)(1+\sqrt{\sqrt{2b}\over l+1}r),
\end{equation}
while for $a=\sqrt{\sqrt{2b}(l+1)}$, i.e. an attractive Coulomb term, we have a first-excited state (one-node):
\begin{equation}\label{eq26}
\psi_1(r)=r^{l+1}\exp\left(-\sqrt{b\over 2}~ r^2\right)\left(1-\sqrt{\sqrt{2b}\over l+1}r\right).
\end{equation}

\noindent In table (\ref{table:nonlin}), we report the first few exact solutions along with the conditions on the potential parameters. Note, the subscripts on the polynomial solutions $f_i(r)$ refer to the possible number of nodes $n$ in the wave function.

\begin{table}[h] \caption{Conditions for Exact Solutions, here $E_{nl}=(n+l+{3\over 2})\sqrt{2b}$ } % title of Table
\centering % used for centering table
\begin{tabular}{l l} % centered columns (4 columns)
\hline\hline %inserts double horizontal lines
$n$ &  $f_n(r)$\\ [0.5ex] % inserts table %heading
\hline % inserts single horizontal line
0   & $f_0(r)=1$\\ % inserting body of the table
& \\
~  & $a=0$\\ % inserting body of the table
\hline
1   & $f_{{a<0, n=0}\atop {a>0, n=1}}(r)=1-{a\over l+1}r$\\ % inserting body of the table
& \\
& $a^2-\sqrt{2b}(l+1)=0$\\
\hline
2  & $f_{{{a=0, n=0}\atop {a<0, n=0}}\atop {a>0, n=2}}(r)=1-{a\over l+1}r+{\sqrt{2b}\over l+1}r^2$\\
&\\
& $a(a^2-\sqrt{2b}(5+4l))=0$ \\
\hline
3   & $f_{{{a>0, n=1, 2}\atop {a<0, n=0,1}}}(r)=1-{a\over l+1}r+{a^2-3\sqrt{2b}(l+1)\over (l+1)(2l+3)}r^2-{1\over 3}{a(a^2-\sqrt{2b}(7l+9))\over (l+1)(l+2)(2l+3)}r^3$\\
& \\
& $a^4-5\sqrt{2b}(3+2l)a^2+18b(2+l)(1+l)=0$\\
\hline
\hline
\end{tabular}
\label{table:nonlin}
\end{table}
\vskip 0.1true in
\noindent For arbitrary values of $a$ and $b$ that do not satisfy the conditions (\ref{eq20}) and (\ref{eq21}), we may use the asymptotic iteration method \cite{hall1} that can be summarized by the following theorem (for details, see \cite{hall1}, section V Theorem 1, and \cite{ciftci2}, equations (2.13)-(2.14)):
\vskip0.1true in
\noindent{\bf Theorem 3:} \emph{Given $\lambda_0\equiv\lambda _{0}(x)$ and $s_0\equiv s_{0}(x)$ in $C^{\infty }$, the differential equation
\begin{equation}\label{eq27}
y''=\lambda_0(x)y'+s_0(x)y
\end{equation}
 has a general solution%
\begin{equation}\label{eq28}
y=\exp \left( -\int\limits^{x}\alpha (t)dt\right) \left[ C_{2}+C_{1}\int%
\limits^{x}\exp \left( \int\limits^{t}\left( \lambda _{0}(\tau )+2\alpha
(\tau )\right) d\tau \right) dt\right]
\end{equation}%
if for some $n>0$
\begin{equation}\label{eq29}
\frac{s_{n}}{\lambda _{n}}=\frac{s_{n-1}}{\lambda _{n-1}}=\alpha (x)
,\quad\mbox{or}\quad\delta _{n}(x)=\lambda _{n}s_{n-1}-\lambda _{n-1}s_{n}=0,
\end{equation}%
where, for $n\geq 1$,
\begin{align}\label{eq30}
\lambda _{n}&=\lambda _{n-1}^{\prime }+s_{n-1}+\lambda_{0}\lambda _{n},\notag \\
 s_{n}&=s_{n-1}^{\prime }+s_{0}\lambda _{n}.
\end{align}
}
\vskip0.1true in
\noindent Thus, for Eq.(\ref{eq16}), with $\lambda_0(r)$ and $s_0(r)$ given by
\begin{equation}\label{eq31}
\left\{ \begin{array}{l}
 \lambda_0(r)=-{1\over r} \left(-2 r^2\sqrt{2b}+2(l+1)\right), \\ \\
  s_0(r)=-{1\over r} \left[\left(2E-(2l+3)\sqrt{2b}\right)r+2a\right],
       \end{array} \right.
\end{equation}
the asymptotic iteration sequence $\lambda_n(x)$ and $s_n(x)$
can be calculated iteratively using (\ref{eq30}). The energy eigenvalues $E\equiv E_{nl}$
of Eq.(\ref{eq16}) can be obtained as roots of the termination condition (\ref{eq29}). According to the asymptotic iteration method (AIM), in particular the study of Brodie \emph{et al} \cite{cham}, unless the differential equation is exactly solvable, the termination
condition (\ref{eq29}) produces for each iteration an expression that depends on both $r$ and $E$ (for given values of the parameters $a$, $b$ and $l$).  In such a case, one faces the problem of finding the best possible starting value $r = r_0$ that stabilizes the AIM process  \cite{cham}. For our problem, we find that the starting value of $r_0=4$ is sufficient to utilize AIM without much worry about the best possible value of $r_0$. For small values of $a$, where the wavefunction is spread out, we may increase $r_0>4$. In Table \ref{table:num}, we report our numerical results, using AIM, for energies $E_{nl}$ for the attractive ($a=1$) and repulsive ($a=-1$) Coulomb term plus the harmonic-oscillator potential.  The numerical computations in the present work were done using Maple version 13 running on an IBM architecture personal computer where we used a high-precision environment. In
order to accelerate our computation we have written our own code for root-finding algorithm using a bisection method, instead of using the default procedure `{\bf Solve}' of \emph{Maple 13}. The numerical results reported in Table \ref{table:num} are accurate to the number of decimals reported. The subscript $N$ refers to the number of iterations used by AIM.

\begin{table}[h] \caption{Eigenvalues $E_{nl}$ for $V(r)=-a/r+br^2$, where $b=0.5$, $a=\pm 1$ and different $n$ and $l$. The subscript $N$ refer to the number of iteration used by AIM.} % title of Table
\centering % used for centering table
\begin{tabular}{|c|c|p{2.5in}|l|c|p{2.5in}|}
\hline
\multicolumn{6}{|c|}{$b=0.5$}\\ \hline
\multicolumn{3}{|c|}{$a=1$}&\multicolumn{3}{|c|}{$a=-1$}\\ \hline
$n$&$l$&$E_{nl}$&$n$&$l$&$E_{nl}$ \\ \hline
$1$&$0$&$2.500~000~000~000~000~000_{N=3,exact}$&$0$&$0$&$2.500~000~000~000~000~000_{N=3,exact}$\\ \hline
$~$&$1$&$3.801~929~609~626~278~046_{N=80}$&$~$&$1$&$3.219~314~119~830~611~360_{N=74}$\\ \hline
$~$&$2$&$4.930~673~420~047~524~772_{N=72}$&$~$&$2$&$4.087~227~795~734~562~981_{N=67}$\\ \hline
$~$&$3$&$6.006~537~298~710~828~780_{N=65}$&$~$&$3$&$5.007~681~882~732~318~957_{N=61}$\\ \hline
$~$&$4$&$7.058~140~776~824~529~475_{N=60}$&$~$&$4$&$5.953~327~675~284~371~524_{N=56}$\\ \hline
\hline
$0$&$0$&$0.179~668~484~653~553~873_{N=97}$&$1$&$0$&$4.380~233~836~413~610~273_{N=97}$\\ \hline
$1$&$~$&$2.500~000~000~000~000~000_{N=3,Exact}$&$2$&$~$&$6.301~066~353~339~463~595_{N=67}$\\ \hline
$2$&$~$&$4.631~952~408~873~053~214_{N=72}$&$3$&$~$&$8.243~517~978~923~477~298_{N=67}$\\ \hline
$3$&$~$&$6.712~595~725~661~429~760_{N=70}$&$4$&$~$&$10.199~062~810~923~865~963_{N=65}$\\ \hline
$4$&$~$&$8.769~519~600~328~899~714_{N=69}$&$5$&$~$&$12.163~259~523~048~320~928_{N=64}$\\ \hline
\hline
\end{tabular}
\label{table:num}
\end{table}

%%%%%%%%%%%%%%%%%%%%%%%%%%%%%%%%%%%%%%%%%%%%%%%%%%%%%%%%%%%%%%%%%%%%%
\section{Exact solutions for the spherically confined $V(r)$}\label{spec}
%%%%%%%%%%%%%%%%%%%%%%%%%%%%%%%%%%%%%%%%%%%%%%%%%%%%%%%%%%%%%%%%%%%%%
\noindent In this section, we consider the confined case of Coulomb and harmonic oscillator system as described by the radial  Schr\"odinger equation (in atomic units)
\begin{equation}\label{eq32}
-{1\over 2}{d^2\psi(r)\over dr^2}+\left[{l(l+1)\over 2r^2}-{a\over r}+br^2\right]\psi(r)=E\psi(r), \quad 0<r<R, \quad b>0.
\end{equation}
where $l=0,1,\dots$ is the angular-momentum quantum number and $\psi(0)=\psi(R)=0$. Here, again, the parameter $a$ is allowed to in $\mathbb R$.
Intuitively, we may assume the following ansatz for the wave function
\begin{equation}\label{eq33}
\psi(r)=r^{l+1}(R-r)\exp(-\alpha r^2-\beta r)f(r),
\end{equation}
where $\alpha$ and $\beta$ are parameters to be determine, and $R$ is the radius of confinement. The $R-r$ factor ensures that the wave function will become zero at $r=R$.
Direct substitution of Eq.(\ref{eq33}) into Eq.(\ref{eq32}) yields the following second-order linear differential equation for $f(r)$:
\begin{align}\label{eq34}
f''(r)&=-2\left({l+1\over r}-{1\over R-r}-2\alpha r-\beta\right)f'(r)\notag\\
&-{1\over r(R-r)}\bigg[(2b-4\alpha^2)r^4+(4R\alpha^2-4\beta \alpha-2Rb)r^3+(4R\alpha\beta-2E-\beta^2+4l\alpha+10\alpha)r^2\notag\\
&+(R\beta^2-6R\alpha+2RE-4Rl\alpha+2l\beta+4\beta-2a)r-2(l+1)+2Ra-2R\beta(l+1)\bigg]f(r)
\end{align}
Clearly, from this equation, we have $\alpha=\sqrt{b/2}$ and $\beta=0$, which reveals the domination of the harmonic oscillator term even in the confined case. Consequently, for $f(r)$, we have
\begin{align}\label{eq35}
f''(r)&=-2\left({l+1\over r}-{1\over R-r}-\sqrt{2b}~ r\right)f'(r)\notag\\
&-{1\over r(R-r)}\bigg[(-2E+(2l+5)\sqrt{2b})r^2+(-3R\sqrt{2b}+2RE-2Rl\sqrt{2b}-2a)r-2(l+1)+2Ra\bigg]f(r)
\end{align}
Although, equation (\ref{eq35}) still does not lie within the framework of Theorem~1, we may make use of the following result (\cite{hakan1}, Theorem 6)

\noindent{\bf Theorem 4.} \emph{A necessary condition for the second-order linear differential equation
\begin{align}\label{eq36}
\left(\sum_{i=0}^{k+2}a_{k+2,i}x^{k+2-i}\right)y''&+\left(\sum_{i=0}^{k+1}a_{k+1,i}x^{k+1-i}\right)y'-\left(\sum_{i=0}^{k}\tau_{k,i}x^{k-i}\right)y=0
\end{align}
to have a polynomial solution of degree $n$ is
\begin{align}\label{eq37}
\tau_{k,0}=n(n-1) a_{k+2,0}+na_{k+1,0},\quad k=0,1,2,\dots.
\end{align}
}
\vskip0.1true in
\noindent Thus for Eq.(\ref{eq35}), or, more explicitly, the differential equation
\begin{align}\label{eq38}
r(R-r)f''(r)&+2\left({(l+1)(R-r)}-r-\sqrt{2b}~ r^2R+\sqrt{2b}~ r^3)\right)f'(r)\notag\\
&+\bigg[(-2E+(2l+5)\sqrt{2b})r^2+(-3R\sqrt{2b}+2RE-2Rl\sqrt{2b}-2a)r+2Ra-2(l+1)\bigg]f(r)=0
\end{align}
to have polynomial solutions of the form $f_n(r)=\sum_{k=0}^n a_k x^k$, it is necessary that
\begin{equation}\label{eq39}
E_{nl}=(n+l+{5\over 2})\sqrt{2b}.
\end{equation}
This is an important formula for $E_{nl}$ that can facilitate greatly our computations based on AIM. We note, first, using Eq.(\ref{eq39}) that Eq.(\ref{eq35}) can be reduced to
\begin{align}\label{eq40}
f_n''(r)&=-2\left({l+1\over r}-{1\over R-r}-\sqrt{2b}~ r\right)f_n'(r)\notag\\
&-{1\over r(R-r)}\bigg[-2 n\sqrt{2b}~ r^2+(2R\sqrt{2b}(n+1)-2a) r+2Ra-2(l+1)\bigg]f_n(r).
\end{align}

\begin{itemize}
\item It is then clear from equation (\ref{eq40})
that, for $n=0$, we have
\begin{equation}\label{eq41}
\left\{ \begin{array}{l}
 E_{0l}=(l+{5\over 2})\sqrt{2b},\quad f_0(r)=1, \\ \\
  \psi_{0l}(r)=r^{l+1}(R-r)\exp(-\sqrt{b\over 2}r^2).
       \end{array} \right.
\end{equation}
if the parameters $a$, $b$ and the radius of confinement $R$ are related by
\begin{equation}\label{eq42}
aR={l+1},\quad a^2=(l+1)\sqrt{2b}.
\end{equation}
The wavefunction given by (\ref{eq41}) represent the ground-state eigenfunction is each subspace labeled by the angular momentum quantum number $l$.
\vskip0.1true in
\item For $n=1$, we, easily, find that
\begin{equation}\label{eq43}
\left\{ \begin{array}{l}
 E_{1l}=(l+{7\over 2})\sqrt{2b},\quad f_{0,1}(r)=1+\left({1\over R}-{a\over l+1}\right)r, \\ \\
  \psi_{1l}(r)=r^{l+1}(R-r)\exp(-\sqrt{b\over 2}r^2)\left(1+\left({1\over R}-{a\over l+1}\right)r\right),
       \end{array} \right.
\end{equation}
only if the parameters $a$, $b$ and $R$ are related by
\begin{equation}\label{eq44}
\left\{ \begin{array}{l}
 \sqrt{2b}={a\over R}-{l+1\over R^2}\Rightarrow b={1\over 2}\left({a\over R}-{l+1\over R^2}\right)^2, \\ \\
  a={1\over R}\left(2l+{5\over 2}\pm{1\over 2}\sqrt{4l+5}\right).
       \end{array} \right.
\end{equation}
Or, more explicitly, for $a$ and $b$ expressed in terms of the radius of confinement $R$, as
\begin{equation}\label{eq45}
\left\{ \begin{array}{l}
 a_\pm={1\over R}(2l+ {5\over 2}\pm{1\over 2}\sqrt{4l+5}),\\ \\
  b_\pm={1\over 8R^4}(\pm(3+2l)+\sqrt{4l+5})^2,
       \end{array} \right.\Longrightarrow E={1\over 2}(7+2l)\sqrt{2b}.
\end{equation}
From (\ref{eq43}), for $a>0$, it is clear that either $a<(l+1)/R$ or $a>(l+1)/R$, since, for the case of $a=(l+1)/R,$ we have $b=0$, which is not acceptable from the structure of our wave function (\ref{eq33}) where $b>0$. We further note from (\ref{eq43}) that for $r<R$ to have one node within $(0,R)$, it is necessary that $R>2(l+1)/a>(l+1)/a$. For example, if $a=1, l=0$, then for a one-node state within $(0,R)$, it is required that $R>2$. Thus, let $a=1,l=0,R={5/2+\sqrt{5}/2}$, we have from (\ref{eq45}), $b=1/50$ and $E_{10}=0.700~000~000~000~000$. Note further, if $a=1,l=0$ but $R=5/2-\sqrt{5}/2$, although $1/R<(l+1)/a$, still we do not have any node that lies within $(0,5/2-\sqrt{5}/2)$, since in this case $r=R=5/2-\sqrt{5}/2$. Thus we have, in this case,  a node-less wave function $f_0(r)$ with $E_{00}=0.699~999~999~995~412~275$. This explains the subscript $f_{0,1}$ in (\ref{eq43}).

\item Further, for $n=2$, we can show that
\begin{equation}\label{eq46}
\left\{ \begin{array}{l}
 E_{2l}={1\over 2}(9+2l)\sqrt{2b},\quad\quad f_{0,1,2}(r)=1+\left({1\over R}-{a\over l+1}\right)r-\left({(3\sqrt{2b}(l+1)-a^2)R^2+(aR-l-1)(2l+3)\over R^2(l+1)(2l+3)}\right)r^2,
\\ \\
 \psi_{2l}(r)=r^{l+1}(R-r)\exp(-\sqrt{b\over 2}r^2)\left(1+\left({1\over R}-{a\over l+1}\right)r-\left({(3\sqrt{2b}(l+1)-a^2)R^2+(aR-l-1)(2l+3)\over R^2(l+1)(2l+3)}\right)r^2\right),
       \end{array} \right.
\end{equation}
only if $a$, $b$ and $R$ are related by the two-implicit expressions:
\begin{equation}\label{eq47}
R^3a^3-3R^2(l+2)a^2-R(R^2\sqrt{2b}(7l+9)-3(l+2)(2l+3))a-3(l+2)(l+1)(2l+3-3R^2\sqrt{2b})
\end{equation}
and
\begin{equation}\label{eq48}
R^2a^3-R(R^2\sqrt{2b}+2l+3)a^2+(l+1)(2l+3-3R^2\sqrt{2b})a+6bR^3(l+1)=0.
\end{equation}
We now consider a few examples of these results:
\begin{itemize}
\item If $a=1,l=0$, then (\ref{eq47}) and (\ref{eq48}) yields $\sqrt{2b}=0.76025880213480504582, R=2.0843217092058454961$, and we have the following solution:
\begin{equation}\label{eq49}
\left\{ \begin{array}{l}
 E_{10}=3.421~164~609~606~622~706~2,
\\ \\
 \psi_{10}(r)=r(2.084~321~709~205~845~496~1-r)\exp(-0.380~129~401~067~402~522~91~r^2)\\
\quad\quad\quad\quad\times \left(1-0.520~227~613~816~384~707~07~r-0.676~516~312~440~766~915~00~r^2\right).
       \end{array} \right.
\end{equation}
\item If $\sqrt{2b}=2,l=0$, then (\ref{eq47}) and (\ref{eq48}) yields $a=-1.6219380762368883824, R=1.0232416568868508038$ and we have the following solution:
\begin{equation}\label{eq50}
\left\{ \begin{array}{l}
 E_{00}=9.000~000~000~000~000,
\\ \\
 \psi_{00}(r)=r(1.023~241~656~886~850~803~8-r)\exp(-r^2)\\
\times \left(1+2.599~224~324~572~826~833~6~r+1.417~080~563~127~630~983~0~r^2\right).
       \end{array} \right.
\end{equation}
\end{itemize}

\item For $n=3$, we have
\begin{equation}\label{eq51}
\left\{ \begin{array}{l}
 E_{3l}={1\over 2}(11+2l)\sqrt{2b}\\ \\
f_2(r)=1+\left({1\over R}-{a\over l+1}\right)r-{R(4R\sqrt{2b}(l+1)-a(Ra-3-2l))-(2l+3)(l+1)\over R^2(2l+3)(l+1)}r^2\\
+{1\over 3}{{(-R^3a^3+3R^2(l+2)a^2+R(R^2\sqrt{2b}(10l+13)-3(l+2)(2l+3))a-3(l+2)(l+1)(-2l+4R^2\sqrt{2b}-3))}\over R^3(l+1)(l+2)(2l+3)}r^3
       \end{array} \right.
\end{equation}
where $a$, $b$ and $R$ are related by
\begin{align}\label{eq52}
R^3a^3&-R^2(R^2\sqrt{2b}+3(l+2))a^2-R(R^2\sqrt{2b}(10l+13)-3(l+2)(2l+3))a\\
&+2b(13+10l)R^4+12\sqrt{2b}(l+1)(l+2)R^2-3(2l+3)(l+2)(l+1)=0
\end{align}
and
\begin{align}\label{eq53}
R^4a^4&-2R^3(2l+5)a^3-R^2(R^2\sqrt{2b}(16l+25)-6(2l+5)(l+2))a^2+2R(2l+5)(R^2\sqrt{2b}(10l+13)\notag\\
&-3(l+2)(2l+3))a+6(l+1)(l+2)((2l+5)(2l+3)+4\sqrt{2b}R^2(R^2\sqrt{2b}-5-2l))=0.
\end{align}
\end{itemize}
Similar results can be obtain for higher $n$ (the degree of the polynomial solutions). It is necessary to note that the conditions reported here  are for the mixed potential $V(r)=-a/r+br^2,$ where $a\neq 0,b \neq 0$ (neither coefficient is zero).

\vskip0.1true in
For the arbitrary values of $a,b$ and $R$, not necessarily satisfying the above conditions, we still apply AIM directly to compute the eigenvalues. Similarly to the un-confined case, we start with
\begin{equation}\label{eq54}
\left\{ \begin{array}{l}
 \lambda_0(r)=-2\left({l+1\over r}-{1\over R-r}-\sqrt{2b}~ r\right), \\ \\
  s_0(r)=-{1\over r(R-r)}\bigg[(-2E+(2l+5)\sqrt{2b})r^2+(-3R\sqrt{2b}+2RE-2Rl\sqrt{2b}-2a)r+2Ra-2(l+1)\bigg].
       \end{array} \right.
\end{equation}
The AIM sequence $\lambda_n(x)$ and $s_n(x)$
can be calculated iteratively using (\ref{eq30}). The energy eigenvalues $E\equiv E_{nl}$
of Eq.(\ref{eq38}) are obtained as roots of the termination condition (\ref{eq31}).
Since the differential equation (\ref{eq38}) has two \emph{regular} singular points at $r=0$ and $r=R$, our initial value of $r_0$ can be chosen to be an arbitrary value in $(0,R)$. In table \ref{table:conf}, we reported the eigenvalues computed using AIM for a fixed radius of confinement $R=1$ with $r_0=0.5$ as an initial value to seed the AIM process. In general, the computation of the eigenvalues are fast as illustrated by the small number of iteration $N$ in Tables \ref{table:conf}, \ref{table:fixb} and \ref{table:varyb}. The same procedure can be applied to compute the eigenvalues for arbitrary values of $a$, $b$ and $R$. In Table \ref{table:fixb} we have fixed $a,b$ and allowed $R$ to vary, then we fixed $b,R$ and allowed $a$ to vary. In Table \ref{table:varyb}, we fixed $a,R$ and varied $b$. Our numerical results
in these tables confirm our earlier monotonicity formulas reported in section II.

\begin{table}[h] \caption{Eigenvalues $E_{nl}$ for $V(r)=-a/r+br^2,~~r\in(0,R)$, where $b=0.5$, $a=\pm 1$, $R=1$ and different $n$ and $l$. The subscript $N$ refers to the number of iteration used by AIM.} % title of Table
\centering % used for centering table
\begin{tabular}{|c|c|p{2.1in}|l|c|p{2.5in}|}
\hline
\multicolumn{6}{|c|}{$a=1,b=0.5,R=1$}\\ \hline
$n$&$l$&$E_{nl}$&$n$&$l$&$E_{nl}$ \\ \hline
$0$&$0$&$~2.500~000~000~000~000~000_{N=3,Exact}$&$0$&$0$&$~~2.500~000~000~000~000~000_{N=3,Exact}$\\ \hline
$~$&$1$&$~8.404~448~391~842~929~575_{N=24}$&$1$&$~$&$~16.733~064~961~893~308~967_{N=25}$\\ \hline
$~$&$2$&$15.183~570~193~031~143~001_{N=23}$&$2$&$~$&$~41.029~002~263~262~675~364_{N=33}$\\ \hline
$~$&$3$&$23.137~256~709~545~767~885_{N=24}$&$3$&$~$&$~75.297~038~665~283~580~892_{N=40}$\\ \hline
$~$&$4$&$32.295~207~272~878~341~541_{N=27}$&$4$&$~$&$119.493~804~921~354~632~859_{N=47}$\\ \hline
\multicolumn{6}{|c|}{$a=-1,b=0.5,R=1$}\\ \hline
$n$&$l$&$E_{nl}$&$n$&$l$&$E_{nl}$ \\ \hline
$0$&$0$&$~7.427~602~986~235~605~737_{N=26}$&$0$&$0$&$~~7.427~602~986~235~605~737_{N=26}$\\ \hline
$~$&$1$&$12.118~629~877~542~593~085_{N=24}$&$1$&$~$&$~22.954~866~627~528~634~394_{N=27}$\\ \hline
$~$&$2$&$18.456~796~172~766~948~526_{N=23}$&$2$&$~$&$~48.054~781~032~847~609~425_{N=36}$\\ \hline
$~$&$3$&$26.173~002~039~626~403~748_{N=25}$&$3$&$~$&$~82.897~495~765~909~946~966_{N=41}$\\ \hline
$~$&$4$&$35.179~533~437~869~611~594_{N=28}$&$4$&$~$&$127.540~759~830~804~826~131_{N=48}$\\ \hline
\hline
\end{tabular}
\label{table:conf}
\end{table}

\begin{table}[h] \caption{Eigenvalues $E_{00}$ for $V(r)=-a/r+br^2,~~r\in(0,R)$, where we fixed $b=0.5$ and we allowed $a$ and $R$ to vary. The subscript $N$ refers to the number of iteration used by AIM.} % title of Table
\centering % used for centering table
\begin{tabular}{|c|c|p{2.2in}||c|c|p{2.2in}|}
\hline
\multicolumn{6}{|c|}{$b=0.5$}
\\ \hline
$a$&$R$&$E_{00}$ & $R$&$a$&$E_{00}$\\ \hline
1&$0.1$&$~468.994~438~340~395~273~843_{N=26}$& 1&-10 &$~~~~24.446~394~090~129~924~468_{N=25}$\\
&$0.5$&$~~14.781~525~455~450~240~772_{N=19}$& ~&-5 &$~~~~15.581~919~590~917~726~881_{N=25}$\\
&$1$&$~~~2.500~000~000~000~000~000_{N=3,Exact}$&~& -1& $~~~~~7.427~602~986~235~605~737_{N=26}$ \\
&$2$&$~~~0.281~457~639~408~567~801_{N=44}$&~& ~0& $~~~~~5.075~582~015~226~783~066_{N=26}$ \\
&$3$&$~~~0.180~768~103~642~728~017_{N=66}$&~ & ~1&$~~~~~2.500~000~000~000~000~000_{N=3,Exact}$\\
&$4$&$~~~0.179~669~842~444~710~526_{N=80}$&~ & ~5&$~-12.356~931~301~584~560~963_{N=35}$\\
&$5$&$~~~0.179~668~484~856~687~713_{N=82}$&~& 10&$~-49.984~937~021~677~890~425_{N=43}$\\
\hline
\end{tabular}
\label{table:fixb}
\end{table}

\begin{table}[!h] \caption{Eigenvalues $E_{00}$ for $V(r)=-a/r+br^2,~~r\in(0,R)$, where we fixed $a=R=1$ and allowed $b$ to vary. The subscript $N$ refer to the number of iterations used by AIM.} % title of Table
\centering % used for centering table
\begin{tabular}{|c|p{2.5in}|}
\hline
\multicolumn{2}{|c|}{$R=1,a=1$}
\\ \hline
$b$&$E_{00}$\\ \hline
0.1&$~~2.399~281~395~696~719~214_{N=22}$\\
0.2&$~~2.424~527~479~482~894~839_{N=22}$\\
0.5&$~~2.500~000~000~000~000~000_{N=3}$\\
1.0&$~~2.624~907~458~899~526~414_{N=31}$ \\
2.0&$~~2.871~465~192~314~860~746_{N=35}$\\
5.0&$~~3.585~958~081~033~459~432_{N=41}$\\
10.0&$~~4.698~782~960~476~752~179_{N=47}$\\
\hline
\end{tabular}
\label{table:varyb}
\end{table}

%%%%%%%%%%%%%%%%%%%%%%%%%%%%%%%%%%%%%%%%%%%%%%%%%%%%
\section{Spectral characteristics}
%%%%%%%%%%%%%%%%%%%%%%%%%%%%%%%%%%%%%%%%%%%%%%%%%%%%
\noindent In this section we shall discuss the spectral characteristics
associated with the crossings of the energy levels. We have
employed the generalized pseudo-spectral (GPS) Legendre method
with mapping, which is a fast algorithm that has been tested
extensively and shown to yield the eigenvalues with an accuracy of
twelve digits after the decimal. A more detailed account, with
several applications of GPS, can be found in
\cite{yao93,cecil01,tong01,roy02,sen06,MAS} and
 the references therein.
\vskip0.1true in
\noindent In the present work, we have also verified
 the accuracy of these results, in a few selected cases, by using
 AIM. We shall first consider the case defined by $R \rightarrow \infty, a=1$
 and variable $b$, under which the potential given by Eq.(1) can be regarded as
 representing the hydrogen atom confined by a soft harmonic oscillator
 potential. Starting from the free HA $(b=0)$, the effect of finite
 $b$ is to remove the accidental degeneracy and raise the energy
 levels such that $E(\nu,\ell) > E(\nu, \ell+1) $ (see Ref.~\cite{baum}). As the starting
 $E(\nu,\ell) < 0 $ given by the HA spectrum, there exists a
 critical value of $b= b_c$, corresponding to each level, defined by the
 condition $E(\nu,\ell)= 0 $. The numerical values of $b_c$ are
 found to be rather small, except for the ground state, indicating
 that a weak confinement due to the harmonic potential is
 sufficient to realize the condition that $E(\nu,\ell) > 0 $ for
 all $b > b_c$. In the usual spectroscopic notations the levels
 $$(1s2s2p3s3p3d4s4p4d4f)$$ are defined by the $b_c$ values given respectively by
\[
b_c = (0.32533,0.004831,0.00771,0.00042,0.00051,0.00079,0.00007,0.00008,0.00010,0.00015).
\]
 In Fig.~1, we have displayed the passing of the energy levels corresponding to $4s4p4d4f$
 states through $E=0$ at $b_c$. We know tht the eigenspectrum of free HA is indeed very sensitive to the
 harmonic confinement since it is found numerically that at
 $b=0.000001$, the eigenvalues are already positive, corresponding
 to the states given by $7p,7d,8d,..,7f,8f,..,7g..$. The crossings
 of energy levels can be gauged by the change of ordering from the
 hydrogen-like
\[
(1s2p2s3d3p3s4f4d5g4p4s5f5d6g5p5s6f6d7g6p6s7f7d8g7p8f8d9g9f\dots)
\]
to
\[
\rightarrow (1s2p2s3d3p4f3s4d5g4p5f4s5d6g5p6f5s6d7g6p7f6s7d8g7p8f8d9g9f\dots)
\]
 as the parameters of the potential change along $(a=1,b=0)$
 $\rightarrow$ $(a=1, b=0.001)$ $\rightarrow$ $(a=1, b=0.5)$. It follows
 that the $(3s,4f),~(4s,5f)\dots$ levels defined by $(\nu,\ell)$ and
 $(\nu+1,\ell+3)$ cross at a certain $b$.

\begin{figure}[!h]
\centering
\includegraphics[height=10cm,width=15cm]{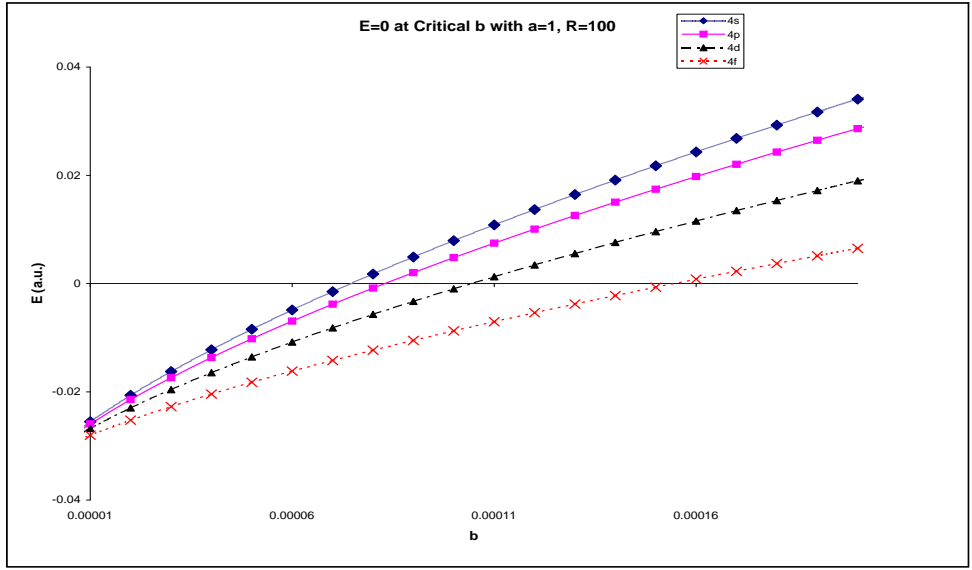}
\caption{The critical b, denoted as $b_c$ in the text at which
$E(\nu,\ell)=0$ are shown for the $4s,4p,4d, 4f$ states. The large
value of $R=100$ corresponds essentially to the free state of the potential in
Eq.(1) with $a=1$.}\label{wf1}
\end{figure}

\noindent In Fig.~2, we have displayed this behavior corresponding to $a=1$.
 This spectral characteristic is similar to that found earlier
 \cite{hssc} for the case of the soft Coulomb potential. Further, the
 eigenvalue $(\nu=5,\ell=4)$ is found to cross  $(\nu-1,\ell-4), (\nu-1,\ell-3)
 (\nu,\ell-4), (\nu,\ell-3) (\nu,\ell-2), (\nu,\ell-1)$ as $b$ changes from $0$ to
 $0.5$.

\begin{figure}[!h]
\centering
\includegraphics[height=10cm,width=15cm]{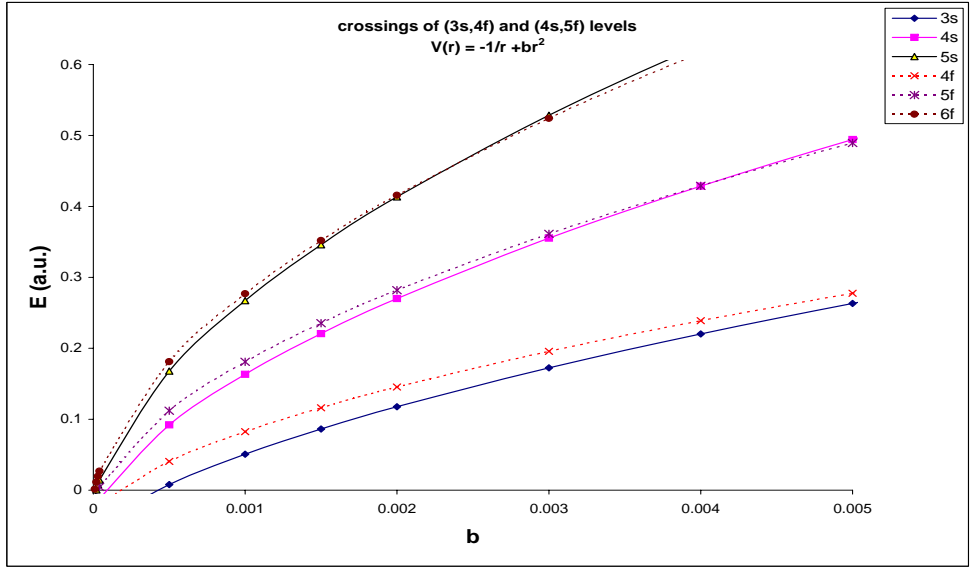}
\caption{The crossings of levels as a function of $b$, corresponding
to the free state of the potential in Eq.(1) with $a=1$. The
levels defined by $(\nu,\ell)$ and
 $(\nu+1,\ell+3)$ are shown. }\label{wf2}
\end{figure}

\vskip0.1true in

\noindent Next, we consider the new spectral characteristics introduced
 when, in addition to the harmonic-oscillator potential term, a second confining
 feature consisting of an impenetrable sphere of finite radius $R$ is introduced.
 Such a potential factor further raises the energy levels as $R$ is diminished,
 $\rightarrow$~0. As a consequence the $b_c$ values get smaller.
 This is depicted in Fig.~3 for the $4s$ and $4p$ states at two
 different values of $R$ of $100$ and $30$ a.u., respectively. The
 former corresponds to the case $R \rightarrow \infty$, i.e. just
 the potential in Eq.(1). Varying $R$ under fixed $a$ yields a
 different level ordering, depending upon the value of $b$, as this
 situation corresponds to two specifically chosen confinement features
imposed on
 the hydrogen-like potential at each point. To illustrate this, we
 consider the case defined by $a=1,\, b=0.5$ and variable $R$. Our
 calculations suggest that the ordering of levels changes from
\[
 (1s2p2s3d3p4f3s4d5g4p5f4s5d6g5p6f5s6d7g6p7f6s7d8g7p8f8d9g9f10g\dots)
\]
to
\[
 \rightarrow
 (1s2p3d2s4f3p5g4d3s5f4p6g5d4s6f5p7g6d5s7f6p8g7d6s8f7p9g8d9f10g\dots)
\]
 as $R$ changes from $\infty \rightarrow 0$. The crossings of levels
 are now observed between the state $(\nu,\ell)$ and
 $(\nu-1,\ell+2)$.

\noindent In Fig.~4, we have shown this feature
 corresponding to the confined $(3s,4d)$ and $(3s,4f)$
 states. Additionally, the $5g$ level is found to fall below $4d$ and
 $3s$ levels, successively, as $R$ decreases. It is evident that the
 imposition of a double confinement effect, mediated through the
 combination of $br^2$ and the boundary at $R$ leads to the crossings among
 a wider set of the states of the hydrogen-like atom, not observed in
 the separate singly confined situations. A possible experimental
 system of embedded atom inside zeolite, fullerine, or liquid helium
 droplets under very strong laser fields could be modelled using
 the doubly confined Coulomb potential as described in this work.

\begin{figure}[htbp]
\centering
\includegraphics[height=10cm,width=15cm]{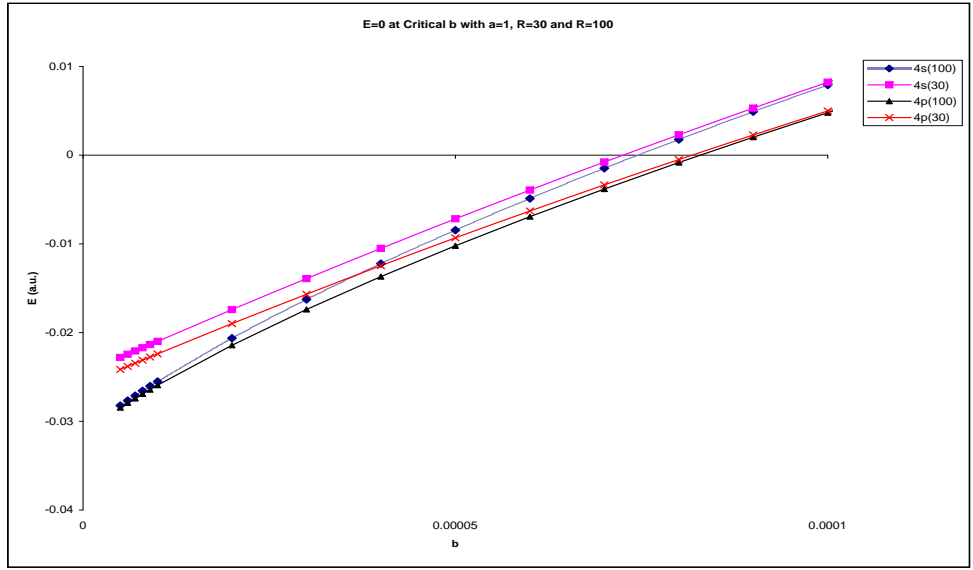}
\caption{The critical $b$, denoted as $b_c$ in the text at which
$E(\nu,\ell)=0$ are shown for the $4s,4p$ states. The value of
$b_c$ decreases as $R$ decreases: specifically, the  essentially free state of the
potential in Eq.(1) with $a=1$ at $R=100$ is confined to a smaller
value of $R=30$. The numbers inside brackets denote
$R$.}\label{wf3}
\end{figure}

\begin{figure}[htbp]
\centering
\includegraphics[height=10cm,width=15cm]{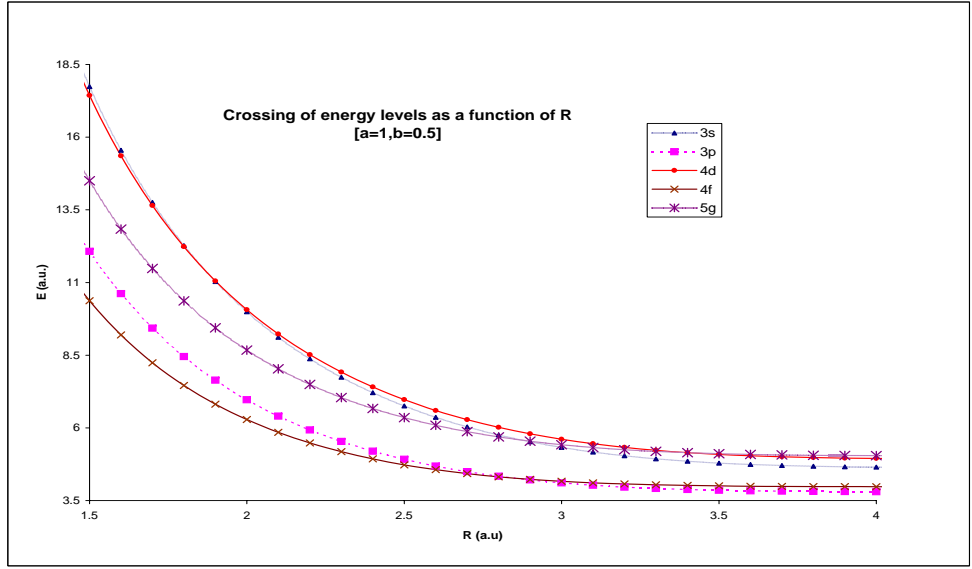}
\caption{The crossings of levels as $R$ is changed as the
potential in Eq.(1) is defined by the values $a=1~ b=0.5$.
Crossings are observed between the state $(\nu,\ell)$ and
$(\nu-1,\ell+2)$ as shown by the $(3s,4d)$ and $(3s,4f)$ levels.
The $5g$ level is shown to cross $4d$ and $3s$ as $R$
decreases.}\label{wf4}
\end{figure}
\newpage
%%%%%%%%%%%%%%%%%%%%%%%%%%%%%%%%%%%%%%%%%%%%%
\section{Conclusion}\label{conc}
%%%%%%%%%%%%%%%%%%%%%%%%%%%%%%%%%%%%%%%%%%%%%
In this study we first consider a very elementary model for an
atom, namely a single particle which moves in a central Coulomb
potential $-a/r$ and obeys quantum mechanics. We then adjoin two
confining features: soft confinement by means of an attractive
oscillator term $br^2,$ and hard confinement produced by
containment inside an impenetrable spherical cavity of radius $R.$
The paper reports on the effects of the confinement parameters
$\{b, R\}$ on the original Coulomb spectrum which, of course, is
given in atomic units by $E = -a^2/(2\nu^2).$ By a combination of
analytical and  numerical techniques, we are able to make
considerable progress in analyzing the spectral characteristics of
this confined atomic model. In future work we plan to undertake a
similar study in which the pure Coulomb term is replaced by a more
physically interesting screened-Coulomb potential, or a soft-core
potential such as $-a/(r+\beta).$ The purpose of this work is to
look at model problems that contain physically interesting
features but are still simple enough to yield to analytical as
well as purely numerical analysis.
% ------------------------------------------------------
\section{Acknowledgments}
% ------------------------------------------------------
\medskip
\noindent Partial financial support of this work under Grant Nos. GP3438 and GP249507 from the
Natural Sciences and Engineering Research Council of Canada
 is gratefully acknowledged by two of us (RLH and NS). KDS is grateful to the Shastri Indo-Canadian
 Institute, Calgary,
 for a partial travel grant. NS and KDS are also grateful for the
 hospitality provided by the Department of Mathematics and Statistics of
Concordia University, where part of this work was carried out.
% --------------------------------------------------------------------------------

\end{document}